\documentclass[prb,twocolumn,showpacs,preprintnumbers,amsmath,amssymb]{revtex4}

\usepackage{graphicx}
\usepackage{dcolumn}
\usepackage{bm}

\begin{document}


\title{Element substitution effect in transition metal oxypnictide Re(O$_{1-x}$F$_x$)TAs (Re=rare earth, T=transition metal)}
\author{G. F. Chen}
\author{Z. Li}
\author{D. Wu}
\author{J. Dong}
\author{G. Li}
\author{W. Z. Hu}
\author{P. Zheng}
\author{J. L. Luo}
\author{N. L. Wang}

\affiliation{Beijing National Laboratory for Condensed Matter
Physics, Institute of Physics, Chinese Academy of Sciences,
Beijing 100080, People¡¯s Republic of China}


\begin{abstract}

Different element substitution effects in transition metal
oxypnictide Re(O$_{1-x}$F$_x$)TAs with Re=La, Ce, Nd, Eu, Gd, Tm,
T=Fe, Ni, Ru, were studied. Similar to the La- or Ce-based
systems, we found that the pure NdOFeAs shows a strong resistivity
anomaly near 145 K, which was ascribed to the spin-density-wave
instability. Electron doping by F increases T$_c$ to about 50 K.
While in the case of Gd, the T$_c$ is reduced below 10 K. The
tetragonal ZrCuSiAs-type structure could not be formed for Eu or
Tm substitution in our preparing process. For Ni-based case,
although both pure and F-doped LaONiAs are superconducting, no
superconductivity was found when La was replaced by Ce in both
cases, instead a ferromagnetic ordering transition was likely to
form at low temperature in F-doped sample. We also synthesized
LaO$_{1-x}$F$_x$RuAs and CeO$_{1-x}$F$_x$RuAs compounds. Metallic
behavior was observed down to 4 K.

\end{abstract}



\maketitle

\section{introduction}

The recent discovery of superconductivity in Fe- and Ni-based
transition metal oxypnictides has generated considerable interest.
The superconductivity was first reported in Fe-based LaOFeP with
transition temperature T$_c\sim$4 K which increases to 7 K with
F$^-$ doping,\cite{Kamihara06} later in Ni-based LaONiP with
T$_c\sim$2 K.\cite{Watanabe} With the replacement of P by As and
partial substitution of O by F in the Fe-based compound to yield
LaO$_{1-x}$F$_x$FeAs, the same group reported that T$_c$ could
rise to 26 K. \cite{Kamihara08} Apparently, element substitution
is an effective way for finding new superconductors with higher
T$_c$ in those families. Guided by this idea, it is indeed found
that the superconducting transition temperature could be
substantially increased in the system of LaO$_{1-x}$F$_x$FeAs when
La was replaced by other rare-earth elements. T$_c$ increases to
41 K by Ce,\cite{Chen1} 43 K by Sm,\cite{XHChen} and 52 K by Pr
replacement.\cite{Zhao} Possibilities for a further increase of
T$_c$ still exist. Here we summarize our recent effort on element
substitution effect in transition metal oxypnictide
ReO$_{1-x}$F$_x$TAs with Re=rare earth elements, T=Fe, Ni, Ru.
Similar to the replacement of La by Ce, Sm, or Pr in
LaO$_{1-x}$F$_x$FeAs, we found that a substitution of Nd for La
increases T$_c$ to about 50 K. While in the case of Gd, the T$_c$
is reduced below 10 K. The same structure phase could not be
formed for Eu or Tm substitution in our preparing process.
Although LaO$_{1-x}$F$_x$NiAs is superconducting, no
superconductivity was found when La was replaced by Ce, instead a
ferromagnetic ordering transition was formed at low temperature.
We also synthesized LaO$_{1-x}$F$_x$RuAs and CeO$_{1-x}$F$_x$RuAs
compounds. Metallic behavior was observed down to 4 K.

\section{Experiment}

\begin{figure}[b]
\includegraphics[width=8cm,clip]{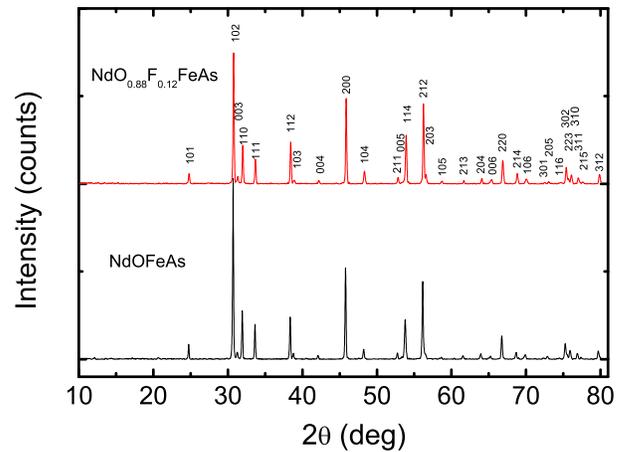}
\caption{(Color online) X-ray powder diffraction patterns of the
pure NdOFeAs and NdO$_{0.88}$F$_{0.12}$FeAs compounds.}
\end{figure}

Different element substitution samples were synthesized by solid
state reaction method using ReAs (Re=Ce, La, Nd, Eu, Gd, Tm),
ReF$_3$, Fe, FeAs, Fe$_2$As, Fe$_2$O$_3$, or rare-earth oxides as
starting materials. ReAs was obtained by reacting Re chips and As
pieces at 500 $^{\circ}C$ for 15 hours and then 850 $^{\circ}C$
for 5 hours. The synthesizing method is similar to those described
in our earlier papers.\cite{Chen1,Chen2,Li,Dong} The raw materials
were thoroughly mixed and pressed into pellets. The pellets were
wrapped with Ta foil and sealed in an evacuate quartz tube. They
were then annealed at 1150 $^{\circ}C$ for 24-50 hours. The
resulting samples were characterized by a powder X-ray
diffraction(XRD) method with Cu K$\alpha$ radiation at room
temperature. The electrical resistivity was measured by a standard
4-probe method. The ac magnetic susceptibility was measured with a
modulation field in the amplitude of 10 Oe and a frequency of 333
Hz. These measurements were preformed in a Physical Property
Measurement System(PPMS) of Quantum Design company.

\section{N\lowercase{d}O$_{1-x}$F$_x$R\lowercase{u}A\lowercase{s} with \lowercase{x}=0 to 0.16}

A series of layered NdO$_{1-x}$F$_x$FeAs compounds with x=0, 0.04,
0.08, 0.12, 0.16 were synthesized. The XRD patterns for the parent
x=0 and x=0.12 compounds are shown in Fig. 1, which could be well
indexed on the basis of tetragonal ZrCuSiAs-type structure with
the space grounp P4/nmm. No impurity phase was identified from the
measurement.

\begin{figure}[t]
\centerline{\includegraphics[width=3.2in]{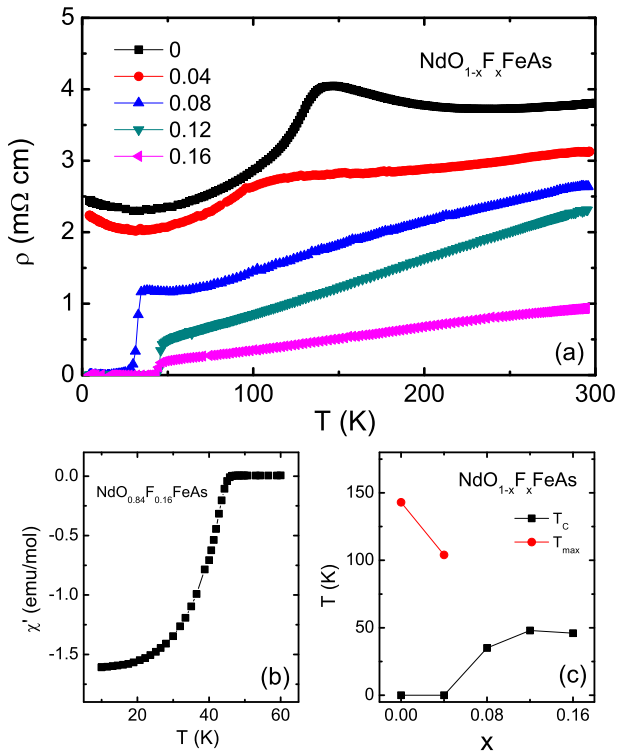}}%
\caption{(Color online) (a) The electrical resistivity vs
temperature for a series of NdO$_{1-x}$F$_{x}$FeAs. (b) Real part
of T-dependent ac magnetic susceptibility for x=0.16. (c) The
phase diagram showing the anomaly (red dots) and superconducting
transition (black square) temperatures as a function of F
content.}
\end{figure}

Figure 2 (a) shows the temperature dependence of the resistivity.
Similar to La- or Ce-based systems, the pure NdOFeAs sample shows
a strong anomaly near 145 K, the resistivity drops steeply below
this temperature. After F-doping, the overall resistivity
decreases and the 145 K anomaly shifts to the lower temperature
and becomes less pronounced. Superconductivity occurs when the
anomaly was removed by electron doping by F substitution for O.
The highest T$_c$ is near 50 K for x=0.12. The bulk
superconductivity in F-doped NdOFeAs is confirmed by ac magnetic
susceptibility measurements. Figure 2 (b) shows the the real part
$\chi'$ of ac susceptibility in a temperature range near Tc for
the x=0.16 sample. Figure 2 (c) is the phase diagram showing the
resistivity anomaly (red cricle) and superconducting transition
(black square) temperatures as a function of F content.

\begin{figure}[b]
\includegraphics[width=8cm,clip]{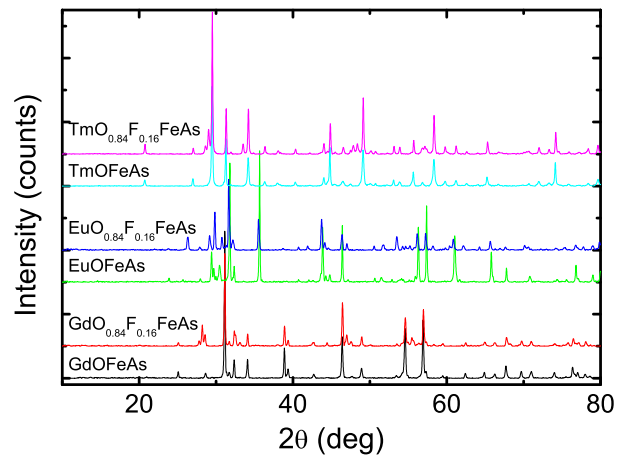}
\caption{(Color online) X-ray powder diffraction patterns for Gd-,
Eu-, and Tm-based ReO$_{1-x}$F$_x$FeAs samples with x=0 and
x=0.16, respectively.}
\end{figure}

We found that the overall behavior of Nd-based series is very
similar to La- or Ce-based series. The resistivity behavior of the
pure NdOFeAs is very similar to that of LaOFeAs. As we elaborated
in our earlier paper, this anomaly is caused by spin-density-wave
(SDW) instability, and a gap opens below the transition
temperature due to the Fermi surface nesting\cite{Dong}. This
strongly suggests that the competing orders are the common feature
for those rare-earth based compounds. High temperature
superconductivity appears near this instability.

\begin{figure}[t]
\centerline{\includegraphics[width=3.2in]{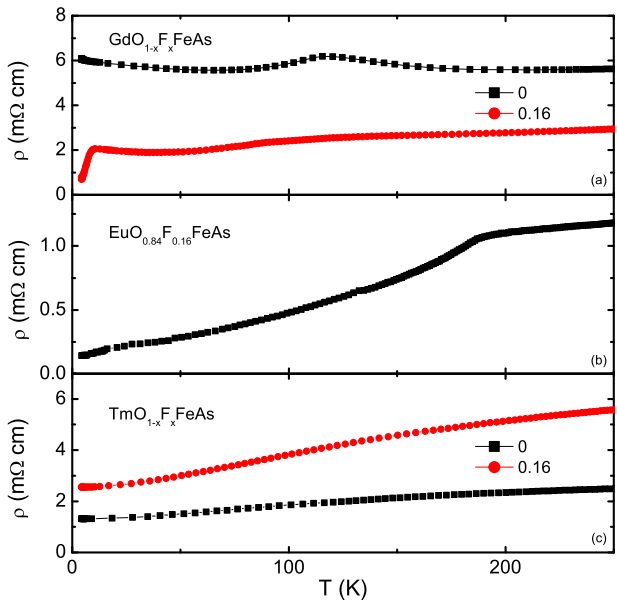}}%
\caption{(Color online) The electrical resistivity vs temperature
for Gd-, Eu-, and Tm-based ReO$_{1-x}$F$_x$FeAs samples with x=0
and x=0.16, respectively.}
\end{figure}

\section{R\lowercase{e}(O$_{1-x}$F$_x$)F\lowercase{e}A\lowercase{s} with R\lowercase{e}=G\lowercase{d}, E\lowercase{u} and T\lowercase{m}}

Besides the Nd-replacement and our earlier work on Ce-based
CeO$_{1-x}$F$_x$FeAs, we also tried to synthesize samples with
other rare-earth element substitutions. Fig. 3 shows the X-ray
diffraction patterns for ReO$_{1-x}$F$_x$FeAs (Re=Gd, Eu, and Tm)
samples with x=0 and x=0.16, respectively. We found that only Gd
replacement could result in almost pure phase. With F-doping,
impurity phase emerges. On the other hand, for Eu and Tm
substitutions, we could not obtain the phase with ZrCuSiAs-type
structure for both pure and F-doped samples. On this basis, we
found that high temperature superconductivity could be easily
realized in light rare-earth element based compounds, but not in
the heavy rare-earth element based systems.

\begin{figure}
\scalebox{0.7}{\includegraphics{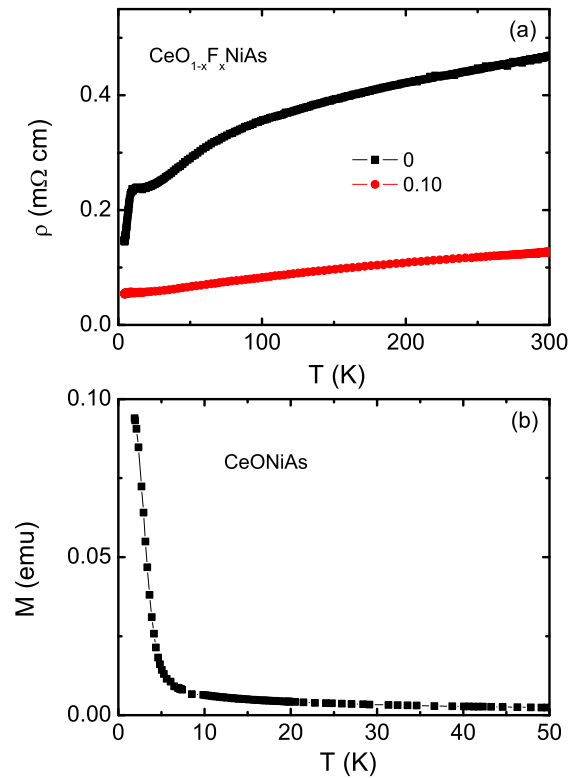}}
\caption{\label{fig:CeONiAs} (Color online) (a) The electrical
resistivity versus temperature T of CeO$_{1-x}$F$_{x}$NiAs with
x=0 and 0.1, respectively. (b) The magnetic moments versus T of
CeONiAs.}
\end{figure}

Figure 4 show the plot of resistivity vs temperature for the above
samples. GdOFeAs shows a similar anomaly as seen in La-, Ce-, or
Nd-based compounds, however this anomaly shifts to lower
temperature. For the x=0.16 F-doped sample, the resistivity shows
a sharp drop below 10 K. Although it could be due to
superoconducting transition, no zero-resistance was obtained. For
the Eu- or Tm-based samples, metallic resistivity was found.
However, as we mentioned above, they do not come from the
structure phase as we expected.

\section{N\lowercase{i}-based C\lowercase{e}O$_{1-x}$F$_x$N\lowercase{i}A\lowercase{s}}

LaONiAs exhibits superconductivity with T$_c \sim$2.75 K. Partial
substitution of oxygen by F increases T$_c$ to about 3.80 K, and
meanwhile dramatically improves the superconducting quality with a
sharp superconducting transition and a high superconducting volume
fraction.\cite{Li} It is of interest to see if a similar
phenomenon could appear in Ce-based CeO$_{1-x}$F$_x$NiAs. Figure
5(a) shows resistivity of CeO$_{1-x}$F$_x$NiAs with x=0 and 0.1.
At high temperatures, resistivity of pure CeONiAs shows similar
T-dependent compared with LaONiAs. Below about 10 K, a sharp
transition is clearly observed. The 10$\%$ F-doping improves the
conductivity while the sharp drop feature at $\sim$ 10 K is
weakened. To distinguish whether this transition is a
superconducting transition or not, we measured the magnetic
moments M below 50 K under a field of 0.1T, as shown in Figure
5(b). We can clearly see a steep increase of M below 10 K,
indicating a ferromagnetic ordering (FM) of magnetic moments. So,
it is most likely that the drop in $\rho$(T) at about 10 K is due
to this FM transition originated from the ordering of Ni$^{2+}$ or
Ce$^{3+}$ moments.

\begin{figure}[]
\centerline{\includegraphics[width=3.2in]{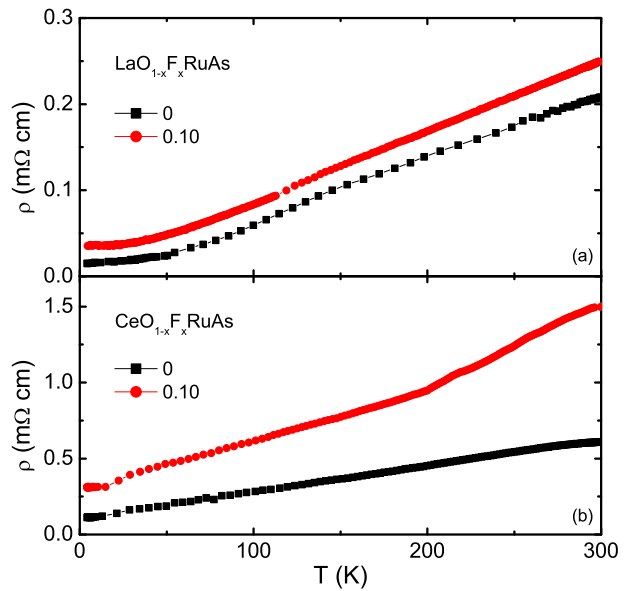}}%
\caption{(a) The electrical resistivity vs temperature for
LaO$_{1-x}$F$_{x}$RuAs (x=0, 0.1). (b) The electrical resistivity
vs temperature for CeO$_{1-x}$F$_{x}$RuAs (x=0, 0.1).}
\end{figure}

\section{R\lowercase{e}O$_{1-x}$F$_{x}$R\lowercase{u}A\lowercase{s} with R\lowercase{e}=L\lowercase{a}, C\lowercase{e}; \lowercase{x}=0, 0.10}

The temperature-dependent resistivity of ReO$_{1-x}$F$_{x}$RuAs
(Re=La, Ce; x=0, 0.10) is shown in Fig. 6. The results clearly
indicate a metallic behavior. No superconductivity was observed
for both pure and doped samples. It is known that the sister
compound CeORuP \cite{Krellner} is a rare case of an ferromagnetic
Kondo lattice. To get more information for the ground state of
CeORuAs, detailed experiments should be done in the future.

\section{Summary}

We studied the element substitution effect in transition metal
oxypnictide ReO$_{1-x}$F$_x$TAs with Re=La, Ce, Nd, Eu, Gd, Tm,
T=Fe, Ni, Ru. Similar to the La- or Ce-based ReO$_{1-x}$F$_x$FeAs
systems, we found that the pure NdOFeAs shows a strong resistivity
anomaly near 145 K, which was ascribed to the spin-density-wave
instability. Electron-doping by F substituting for O increases
T$_c$ to about 50 K. While in the case of Gd, the T$_c$ is reduced
below 10 K. The tetragonal ZrCuSiAs-type structure could not be
formed in Eu or Tm substitutions in our heating process. We
speculate that high temperature superconductivity could be easily
realized in light rare-earth element based compounds, but not in
the heavy rare-earth element based systems. For the Ni-based case,
although both pure and F-doped LaONiAs are superconducting, no
superconductivity was found when La was replaced by Ce in those
compounds, instead a ferromagnetic ordering transition was likely
to form at low temperature. We also synthesized
LaO$_{1-x}$F$_x$RuAs and CeO$_{1-x}$F$_x$RuAs compounds. No
superconductivity is found in both systems.

\begin{acknowledgments}
We acknowledge the support from Y. P. Wang and Li Lu, and Z. X.
Zhao for sharing the preprint of ref.6 before publication. We also
thank H. Chen and S. K. Su for their help in the experiments. This
work is supported by the National Science Foundation of China, the
Knowledge Innovation Project of the Chinese Academy of Sciences,
and the 973 project of the Ministry of Science and Technology of
China.

\end{acknowledgments}


\end{document}